\documentclass[twocolumn]{aastex62}

\newcommand{\feh}{$\mbox{[Fe/H]}$}

\received{October, 2020}
\revised{--}
\accepted{--}

\submitjournal{ApJS}

\shorttitle{Stellar Metallicities of Southern Stars from SkyMapper Photometry}
\shortauthors{Chiti et al.}

\newcommand{\teff}{\ensuremath{T_\mathrm{eff}}}
\newcommand{\logg}{\ensuremath{\log\,g}}

\begin{document}

\title{Stellar Metallicities from SkyMapper Photometry II: \\Precise photometric metallicities of $\sim$280,000 giant stars with [Fe/H] $< -0.75$ in the Milky Way}

\correspondingauthor{Anirudh Chiti}
\email{achiti@mit.edu}

\author[0000-0002-7155-679X]{Anirudh Chiti}
\affil{Department of Physics and Kavli Institute for Astrophysics and Space Research, Massachusetts Institute of Technology, Cambridge, MA 02139, USA}

\author[0000-0002-2139-7145]{Anna Frebel}
\affiliation{Department of Physics and Kavli Institute for Astrophysics and Space Research, Massachusetts Institute of Technology, Cambridge, MA 02139, USA}

\author[0000-0001-9178-3992]{Mohammad K.\ Mardini}
\affiliation{Key Lab of Optical Astronomy, National Astronomical Observatories, Chinese Academy of Sciences, Beijing 100101, China}
\affiliation{Institute of Space Sciences, Shandong University, Weihai 264209, China}

\author{Tatsuya W. Daniel}
\affiliation{Department of Physics, Brown University, Providence, RI 02912, USA}

\author{Xiaowei Ou}
\affiliation{Department of Physics and Kavli Institute for Astrophysics and Space Research, Massachusetts Institute of Technology, Cambridge, MA 02139, USA}

\author{Anastasiia V. Uvarova}
\affiliation{Department of Physics and Kavli Institute for Astrophysics and Space Research, Massachusetts Institute of Technology, Cambridge, MA 02139, USA}

\begin{abstract}

The Milky Way's metal-poor stars are nearby ancient objects that are used to study early chemical evolution and the assembly and structure of the Milky Way. 
Here we present reliable metallicities of $\sim280,000$ stars with $-3.75 \lesssim$ [Fe/H] $\lesssim -0.75$ down to $g=17$ derived using metallicity-sensitive photometry from the second data release (DR2) of the SkyMapper Southern Survey. 
We use the dependency of the flux through the SkyMapper $v$ filter on the strength of the Ca II K absorption features, in tandem with SkyMapper $u,g,i$ photometry, to derive photometric metallicities for these stars.
We find that metallicities derived in this way compare well to metallicities derived in large-scale spectroscopic surveys, and use such comparisons to calibrate and quantify systematics as a function of location, reddening, and color.
We find good agreement with metallicities from the APOGEE, LAMOST, and GALAH surveys, based on a standard deviation of $\sigma\sim0.25$\,dex of the residuals of our photometric metallicities with respect to metallicities from those surveys.
We also compare our derived photometric metallicities to metallicities presented in a number of high-resolution spectroscopic studies to validate the low metallicity end ([Fe/H] $< -2.5$) of our photometric metallicity determinations.
In such comparisons, we find the metallicities of stars with photometric [Fe/H] $< -2.5$ in our catalog show no significant offset and a scatter of $\sigma\sim$0.31\,dex level relative to those in high-resolution work when considering the cooler stars ($g-i > 0.65$) in our sample. 
We also present an expanded catalog containing photometric metallicities of $\sim720,000$ stars as a data table for further exploration of the metal-poor Milky Way.

\end{abstract}

\keywords{Stars: Metal-poor--- Galaxy: abundances}

\section{Introduction} 
\label{sec:intro}

Metal-poor stars\footnote{Defined as $\feh\,\,\le -1\,\text{dex}$, where [Fe/H] = $\log_{10}(N_{\text{Fe}}/N_{\text{H}})_{\star}-\log_{10}(N_{\text{Fe}}/N_{\text{H}})_{\sun}$ \citep{bc+05, fn+15}.} in the Milky Way and their role in exploring the early universe have been studied in great detail for several decades \citep{bps+85, bps+92, c+03, fcn+06, abw+06, smy+17, ltz+18}. 
It has long been known that finding the most metal deficient ([Fe/H] $< -3.0$) of these stars is a challenge because of their extremely rare occurrence rate \citep{bond81}. 
As one illustration, these stars are more likely to be found as part of the Milky Way halo component, but the halo-to-disk star ratio is $\sim10^{-3}$ in the solar neighborhood \citep{bs+80}.
More broadly, the number of stars roughly decreases by a factor of $\sim$10 or more for each dex decrease in metallicity \citep{h+76}. 
For the solar neighborhood, this translates to expectations that one star with $\mbox{[Fe/H]} = -3.0$ may be found among every 65,000 stars, and one star with $\mbox{[Fe/H]} = -3.5$ among 200,000 stars. 
Efficient selection techniques are thus required to make large-scale progress of understanding the old halo component with sufficient statistics. 

Early searches for metal-poor halo stars were based on kinematics (i.e., high proper motions) and were the first systemic discoveries of tens of stars with $\mbox{[Fe/H]} < -3.0$ \citep[e.g.,][]{rn+91}.
Then came the era of low resolution (R$\sim400$) objective-prism spectroscopic surveys of large portions of the sky based on calibrated measurements of the Ca\,II\,K line to obtain metallicities for millions of stars. 
The Southern HK Survey \citep{bps+92} and the Hamburg/ESO Survey \citep{nsf+08, fcn+06} resulted in thousands of candidates with $\mbox{[Fe/H]} < -3.0$. 
However, these candidates required medium-resolution (R$\sim2,000$) follow-up spectra to confirm the metallicity before investing in high-resolution spectroscopic observations to obtain a detailed chemical abundance analysis.
\citet{bc+05} provide a detailed account of all these efforts, including the more recent Northern Sloan Digital Sky Survey (SDSS)\footnote{http://www.sdss.org} and the subsequent SEGUE medium-resolution spectroscopic surveys that extended the reach to fainter stars \citep{yrn+09, ewa+11}. 
For additional details on search and analysis techniques, and how the most metal-poor stars in the halo and dwarf galaxies are utilized to reconstruct the conditions of the early Galaxy, we also refer the reader to \citet{fn+15}.

Promising recent searches for metal-poor stars are based on custom photometric filters designed to facilitate metallicity measurements of millions of stars as part of large sky surveys. 
A metallicity sensitive imaging filter centered near the Ca\,II\,K line that enabled metallicity measurements in the very and even extremely metal-poor regime \citep{bbs+11} was developed for the SkyMapper Telescope and a targeted search for metal-poor stars in the Southern Sky \citep{ksb+07}. 
The Southern Sky Survey has recently provided shallow data of the entire Southern sky (data release DR1.1, \citealt{wol+18}), following data collection in the commissioning phase. 
A number of impressive findings have been achieved, including identification of the most iron-poor stars known in the halo \citep{kbf+14, nbd+19}, large samples of metal-poor stars analyzed with medium \citep{dbm+19} and high-resolution spectroscopy \citep{jkf+15,mda+19}, and a kinematic analysis \citep{cdy+20}. 
In addition, exploration of the bulge revealed a population of extremely metal-poor stars \citep{hca+15}, and spatial maps of the Southern sky as a function of metallicity (down to [Fe/H] $\sim-2.0$) were produced \citep{cwm+19, hcy+19}.

A modified and tuned metallicity-sensitive filter was used to carry out the Pristine survey \citep{smy+17} in the Northern hemisphere with the Canada-France-Hawai'i Telescope. 
Samples of very and extremely metal-poor stars have also been identified in great numbers and followed-up with medium and high-resolution spectroscopy in the halo \citep{sab+18} and the bulge \citep{asm+20}, in addition to the discovery of an ultra metal-poor ([Fe/H] $< -4.0$) star \citep{sab+18}.
The large number of stars with photometric metallicities was also able to probe the extremely metal-poor regime of the metallicity distribution function of the Milky Way \citep{ksm+20}. 
In addition to these targeted surveys for metal-poor stars, stellar photometric metallicities have also been determined from broadband SDSS data of the Northern hemisphere \citep{isj+08, ab+20}.

Recently, Data Release 2 (DR2) of the SkyMapper Southern Sky Survey was released \citep{owb+20}, containing $\sim500$ million astrophysical sources and increasing the imaging depth significantly compared to the earlier data release (DR1.1). 
Chiti et al. (ApJL accepted) presented a metallicity map of the Southern sky with a metallicity resolution down to $\mbox{[Fe/H]}\sim -3.3$, based on sample of $\sim$111,000 stars with photometric metallicities determined by applying techniques in \citet{cfj+20} on SkyMapper DR2 data. 
Here we present the full catalog of metallicities of stars from SkyMapper DR2 from which that sample was selected, composed of giant stars with $g < 17$ and $-3.75<\mbox{[Fe/H]}<-0.75$ for which we achieved metallicities with uncertainty $\lesssim 0.75$ dex.
Our full catalog contains $\sim720,000$ stars, of which $\sim280,000$ have reliable metallicities after applying checks on the evolutionary status and metallicities of these stars using \textit{Gaia} EDR3 data \citep{gaia+16, gaia+20}.

\begin{figure*}[t!]
\includegraphics[width =\textwidth]{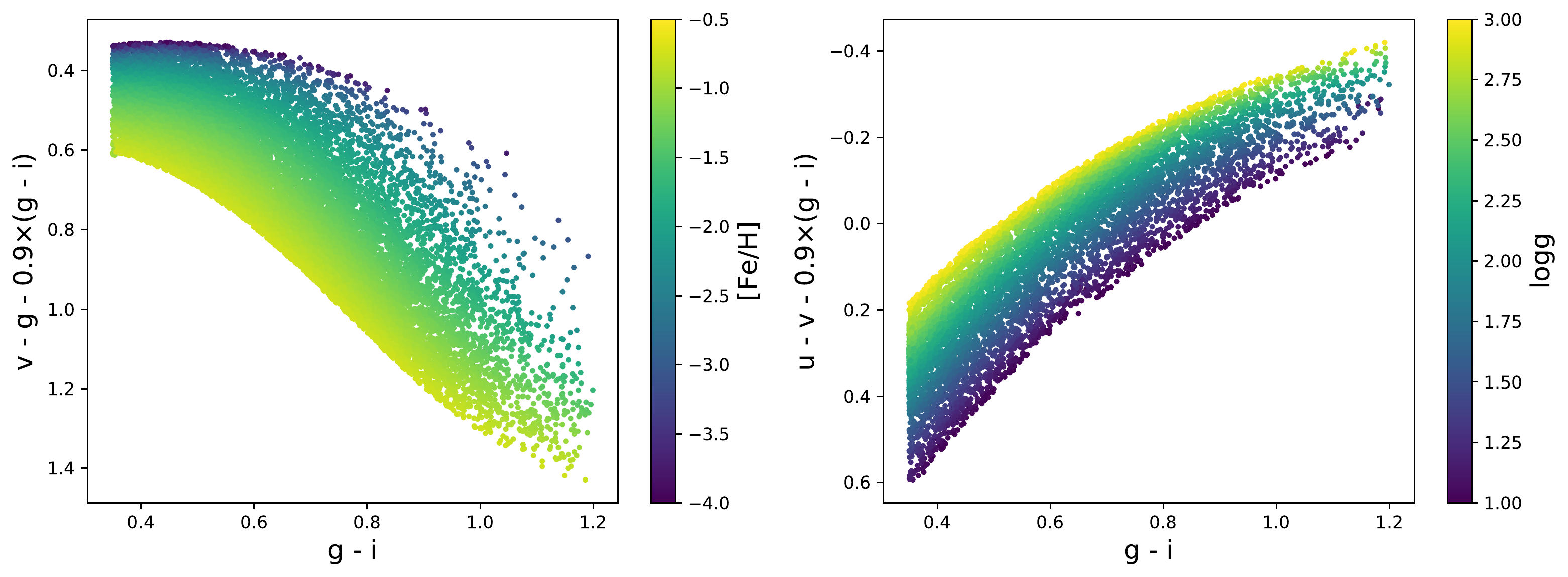}
\caption{Left: Metallicity-sensitive color-color plot that was used to derive photometric metallicities (see Section~\ref{sec:derive}), where each point corresponds to a star in SkyMapper DR2. Right: Surface gravity-sensitive color-color plot, where each point also corresponds to a star in SkyMapper DR2.  }
\label{fig:photometric_metallicity}
\end{figure*}

\section{Methods} 
\label{sec:methods}

Here we outline the steps involved in deriving and compiling photometric metallicities of stars in the second data release (DR2) of the SkyMapper Southern Sky Survey \citep{owb+19}.
We first describe how we compiled the SkyMapper photometric catalog, and then we outline our derivation of photometric metallicities.

\subsection{Compilation of photometric catalog}
\label{sec:compile}

We queried the entire SkyMapper catalog to retrieve photometry of all sources brighter than $g=17$ using the Virtual Observatory Access (TAP) protocol\footnote{http://skymapper.anu.edu.au/how-to-access/}\footnote{http://api.skymapper.nci.org.au/public/tap/} with the \texttt{TapPlus} class in the \texttt{astroquery} python package \citep{gsb+19}.
To ease computing, individual queries were performed on 15$^{\circ}$ by 15$^{\circ}$ regions that, when combined, tiled the full southern hemisphere. 
The photometry was then de-reddened using maps from \citet{sfd+98} with bandpass absorption coefficients listed on the SkyMapper website\footnote{http://skymapper.anu.edu.au/filter-transformations/}.
In all analysis, we used the petrosian magnitudes, which were denoted in the catalog by e.g., \texttt{g\_petro}.
We also only retained objects with the keys \texttt{flags=0} and \texttt{class\_star$>$0.9} in the SkyMapper DR2 catalog.
This initial cut ensured that our sample was composed of stars with no obvious issues e.g., blending affecting their photometry.

\subsection{Initial derivation of photometric metallicities and stellar parameters}
\label{sec:derive}

The photometric metallicities of our stars were derived using the methods in \citet{cfj+20}, which we briefly outline here.
First, we generated a grid of flux calibrated synthetic spectra over a range of stellar parameters (see Table~\ref{tab:grid}) using the Turbospectrum software \citep{ap+98, p+12}, MARCS model atmospheres \citep{gee+08}, and a linelist from VALD database \citep{pkr+95, rpk+15} supplemented by carbon molecular lines from  \citet{bbs+13, brw+14, mpv+14, rbb+14, slr+14}.
The [$\alpha$/Fe] values was set to match the ``standard" MARCS model atmospheres, in which [$\alpha$/Fe] = 0.4 when [Fe/H] $<$ $-$1.0, [$\alpha$/Fe] = 0 when [Fe/H] = 0, and [$\alpha$/Fe] decreases linearly between the two values when $-1$ $<$ [Fe/H] $<$ 0. This behavior matches the general [$\alpha$/Fe] trend in the Milky Way halo.
Then, we derived a grid of synthetic photometry by calculating the expected flux that each of our synthetic spectra would produce through each of the SkyMapper filters \citep{bbs+11}.


\begin{deluxetable}{cccc} 
\tablecolumns{3}
\tablewidth{\columnwidth}
\tablecaption{\label{tab:grid} Stellar parameters of grid of synthetic spectra}
\tablehead{   
  \colhead{Parameter} &
  \colhead{Minimum} &
  \colhead{Maximum} &
  \colhead{Step}
}
\startdata
$\lambda$ & 3000\,\AA & 9000\,\AA & 0.01\,\AA\\
$T_{\text{eff}}$ & 4000\,K & 5700\,K & 100\,K\\
log $g$ & 1.0 & 3.0 & 0.5\\
$[$Fe/H$]$ & $-$4.0 & $-$0.5 & 0.5\\
\enddata

\end{deluxetable}


To derive photometric metallicities, we then  matched the magnitudes in SkyMapper DR2 photometry to those in our grid of synthetic photometry and found the associated metallicity.
Specifically, stars with distinct metallicities separate in a well-behaved manner in the color-color plot $v - g \times 0.9\,(g - i)$ vs $g-i$ (see Figure~\ref{fig:photometric_metallicity}).
Consequently, we overlaid our synthetic photometry on this plot and interpolated between the contours of constant metallicity using the \texttt{scipy.interpolate.griddata} to map each location in that plot to a metallicity.
Then, simply by overlaying the observed SkyMapper photometry onto this plot, we were able to determine metallicities.

We note that the metallicity contours in the color-color plot $v - g \times 0.9\,(g - i)$ vs $g-i$ are dependent on surface gravity (see Figure 3 in \citealt{cfj+20}), so we iteratively determined photometric metallicities to account for this fact.
As such, we first derived photometric metallicities assuming $\log g$ = 2.0.
Then, we derived photometric surface gravities using the fact that stars with  distinct surface gravities separate in a well-behaved manner in the color-color plot $u - v \times 0.9\,(g - i)$ vs $g-i$ (see Figure~\ref{fig:photometric_metallicity}).
As such, we could simply replicate the procedure described for photometric metallicities to derive photometric surface gravities, using the first-pass metallicity estimates as inputs to generate contours for each star.
Then, we used these photometric surface gravity measurements as inputs to adjust the photometric metallicity contours and derive updated, final photometric metallicities.

We note that stars may have stellar parameters or metallicities beyond our grid of synthetic spectra, and we outline how we account for those situations here.
In the case of metallicity, stars with SkyMapper photometry consistent with [Fe/H] $> -0.5$ or [Fe/H] $< -4.0$ would be beyond the edge of our grid.
To avoid the spurious presence of such stars in our catalog, we exclude stars with photometric [Fe/H] $> - 0.75$ or photometric [Fe/H] $< -3.75$. 
We exclude all stars with {\logg} beyond the upper edge of our grid ({\logg} = 3.0), but do retain stars with {\logg} below the lower range of our grid ({\logg} = 1.0) as the effect on the photometric metallicity contours from {\logg} is not hugely significant below that value.
Finally, we account for stars with {\teff} beyond the range of our grid by excluding stars with $g-i < 0.35$ and $g-i > 1.2$.

We derived uncertainties on our photometric metallicities by adding sources of random uncertainty and an estimate of the systematic uncertainty in quadrature. 
We derived random uncertainties by varying the $v - g \times 0.9\,(g - i)$ and $g-i$ colors for each star by propagating the photometric uncertainties in SkyMapper DR2.
After each term was varied, photometric metallicities were re-derived and the differences between the re-derived metallicity and the original metallicity were added in quadrature to derive the total random uncertainty. 
The intrinsic systematic uncertainty from this method was taken as 0.16\,dex, following \citet{cfj+20}.
The random and systematic uncertainties were then added in quadrature for each star to derive a final uncertainty on the photometric metallicity.

\begin{figure}[t!]
\includegraphics[width =\columnwidth]{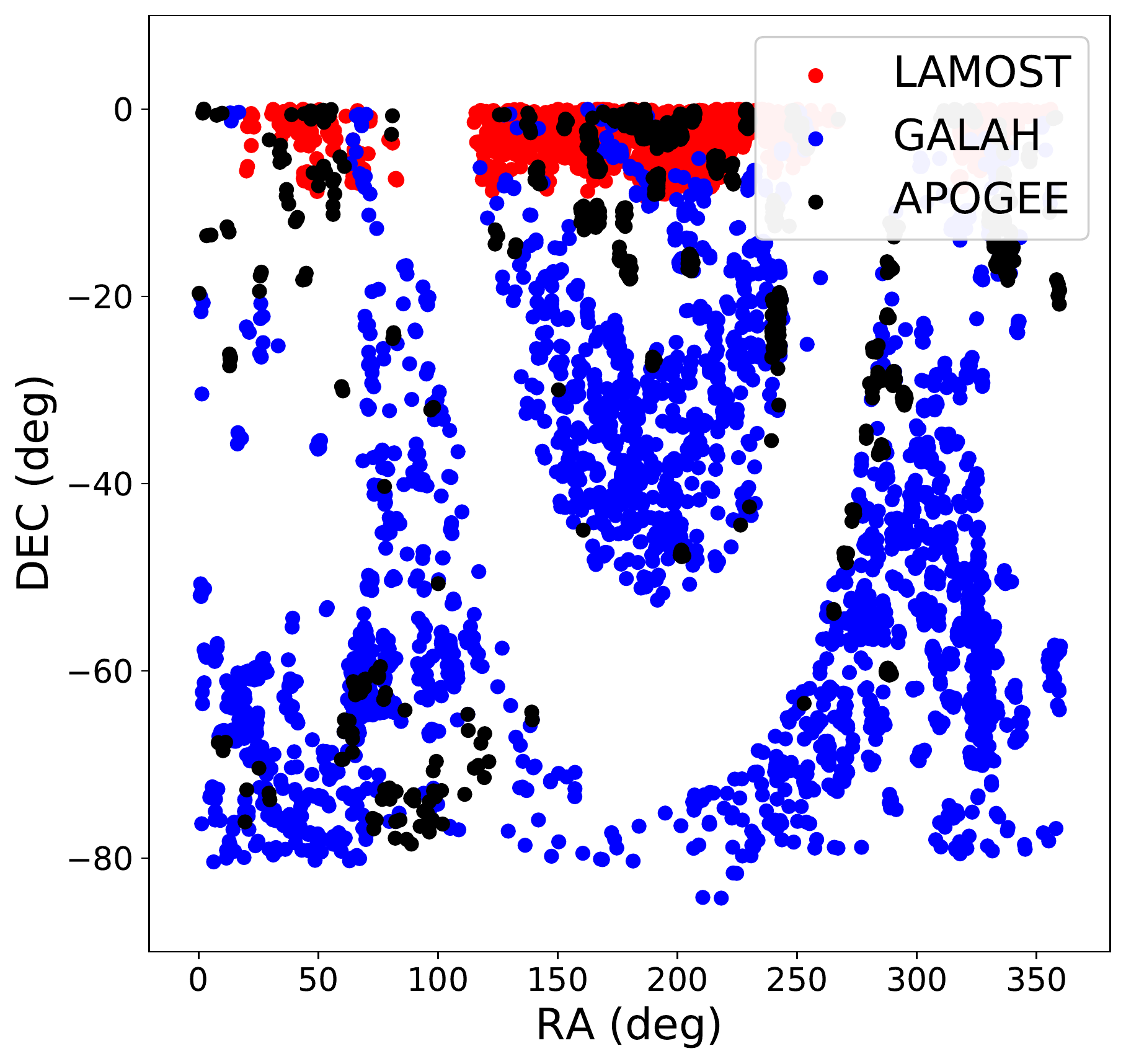}
\caption{Spatial location of stars with SkyMapper photometric metallicities that also have high-quality metallicities (see Section~\ref{sec:surveys} for a description) from LAMOST DR6 (red), GALAH DR3 (blue), and APOGEE DR16 (black).}
\label{fig:spatial_coverage}
\end{figure}

\begin{figure*}[t!]
\includegraphics[width =\textwidth]{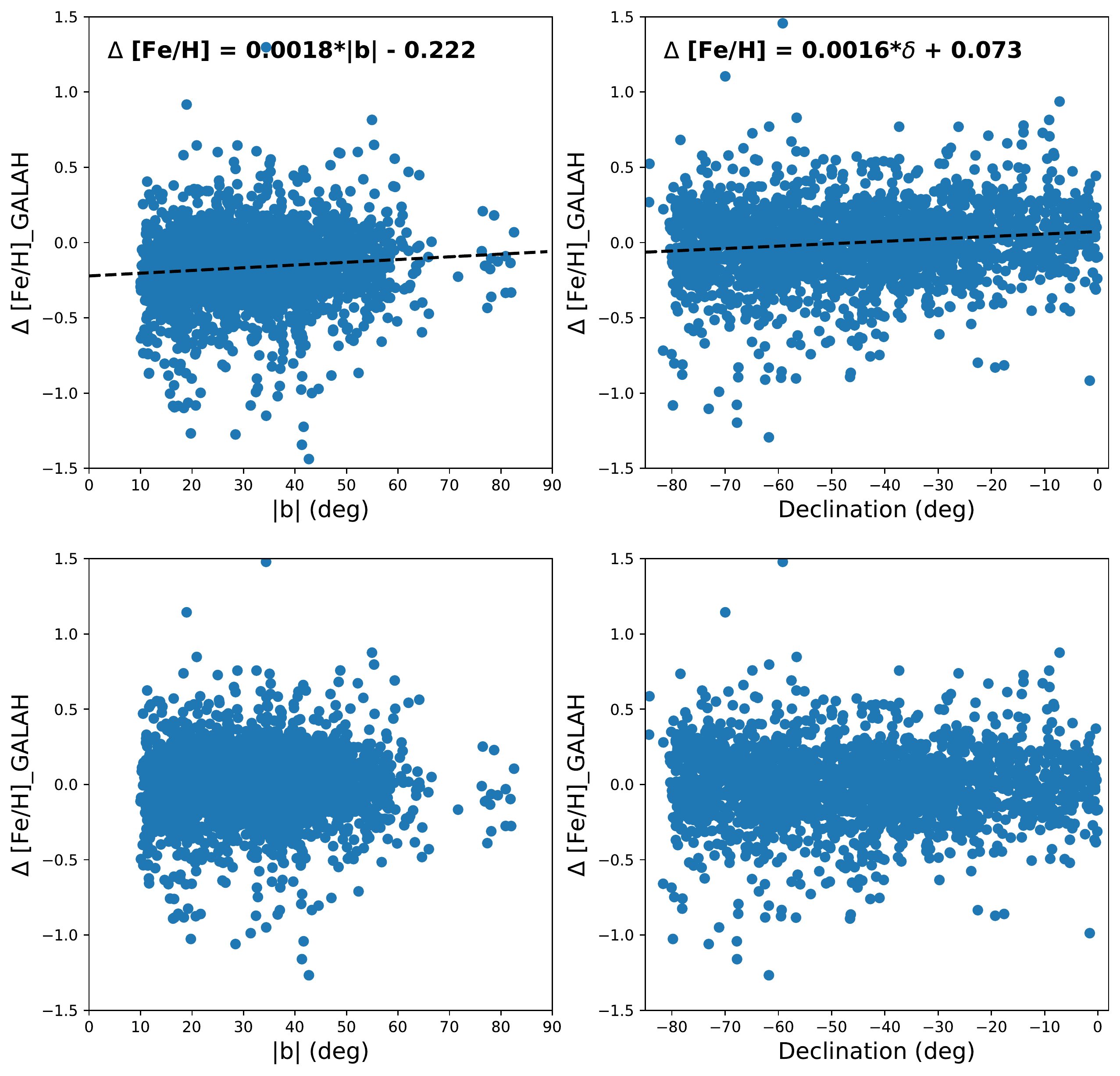}
\caption{Top left: Residuals of the initial photometric metallicities derived in Section~\ref{sec:derive} with respect to metallicities of stars in GALAH DR3 as a function of galactic latitude. 
A line is fit to the residuals and the trend is subtracted from the initial photometric metallicity determination to account for the spatial variation of the metallicity.
Top right: Same as top left, but shown as a function of declination. 
Bottom left: Residuals of the photometric metallicities after removing the trend as a function of galactic latitude. No further trends are apparent.
Bottom right: Same as bottom left, but shown as a function of declination.}
\label{fig:spatial_offset}
\end{figure*}

\subsection{Refinement of stellar metallicities through comparison to large-scale sky surveys}
\label{sec:refinement}

\begin{figure*}[t!]
\includegraphics[width =\textwidth]{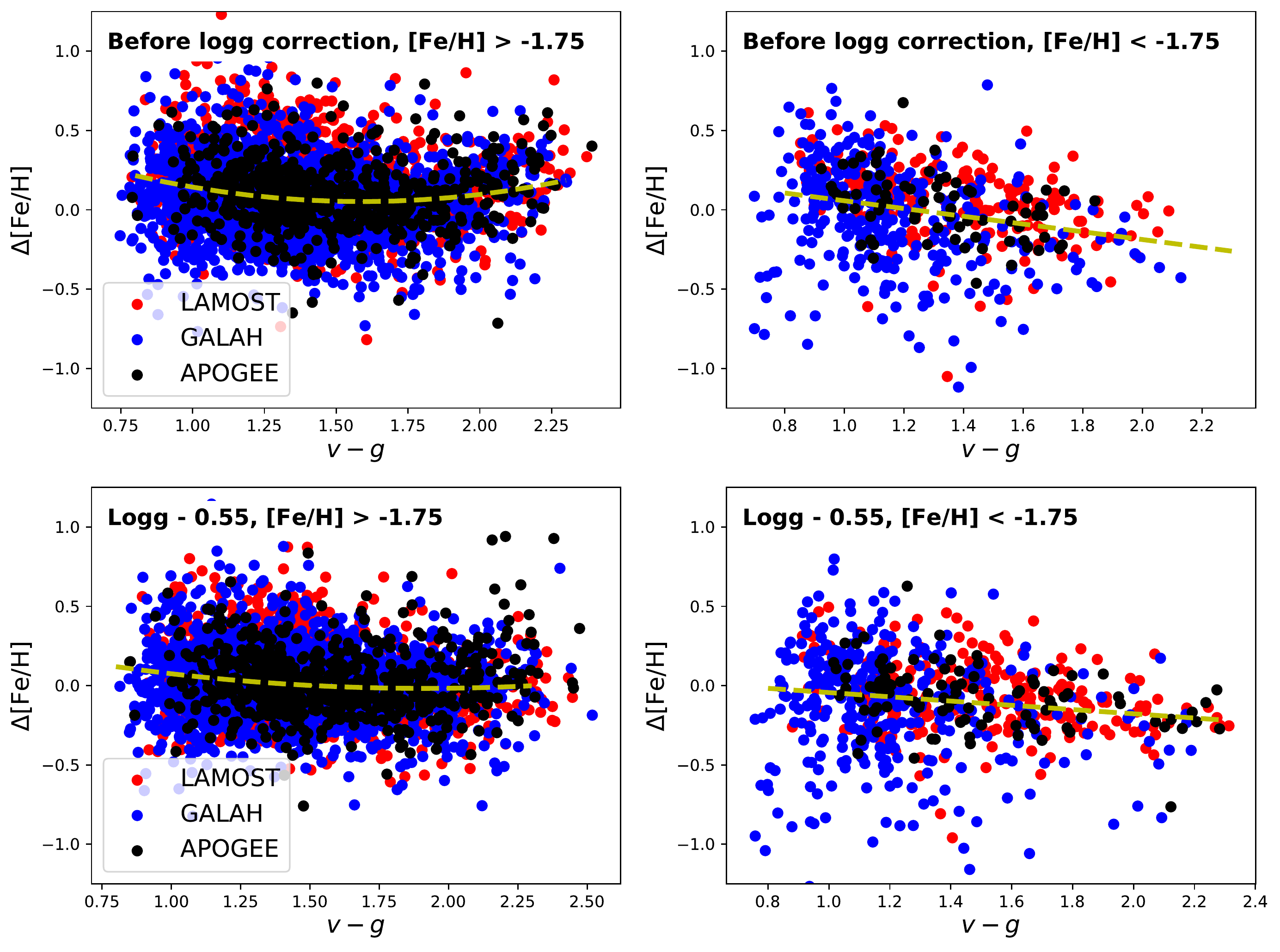}
\caption{Top left: Residuals of our initial photometric metallicities derived in Section~\ref{sec:derive} with respect to high-quality (see Section~\ref{sec:surveys} for a full description) metallicities of stars in LAMOST DR6 (red), GALAH DR3 (blue), and APOGEE DR16 (black) as a function of the $v-g$ color for stars with photometric metallicities [Fe/H] $> -1.75$. 
A clear quadratic trend is apparently in the residuals.
Note that the residuals with respect to each survey have been slightly adjusted to account for zero-point offsets in metallicities.
Top right: Same as left, but residuals are shown for stars with photometric [Fe/H] $< -1.75$ for which a slightly different trend is apparent.
Bottom left and right: Same as top panels, but using photometric metallicities that were calculated after decreasing the photometric $\log\,g$ by 0.55\,dex to bring our surface gravity scale in agreement with that in \citet{erf+20} and Ou et al., in prep.
Only negligible trends in the residual metallicities exist after the surface gravity correction.}
\label{fig:vg_processing}
\end{figure*}

Due to the large size and broad spatial coverage of the SkyMapper catalog, it was necessary to test for systematic effects in our metallicities as a function of sky location, color, and reddening. 
To accomplish this, we compared our photometric metallicities to metallicities derived in three large spectroscopic surveys: APOGEE DR16  \citep{ewa+11, bba+17, msf+17}, GALAH DR3 \citep{galahdr3}, and LAMOST DR6 \citep{lamost_1ref,lamost_2ref,lamost_3ref}.
 A full description of the cross-matching procedure and quality criteria that were applied to compile a comparison sample from these large spectroscopic surveys is detailed in Section~\ref{sec:surveys}.
 For the investigations in this section, we only included stars from those surveys in the metallicity range of our sample ($-3.75 < $[Fe/H] $< -0.75$) to ensure a consistent comparison sample.

We first checked for any trends in the residuals of our photometric metallicities with respect to the metallicities from GALAH DR3 as a function of right ascension, declination, galactic longitude ($l$), and galactic latitude ($b$).
The GALAH survey was chosen for the comparison sample due to its broad overlap in spatial coverage and magnitude range with our catalog of photometric metallicities (see Figure~\ref{fig:spatial_coverage}).
We find no trends with respect to right ascension and galactic longitude, but find statistically significant trends as a function of declination and galactic latitude (see Figure~\ref{fig:spatial_offset}).
The trend with respect to declination likely arises from residual effects on the photometry from the airmass. 
Similarly, the trend with respect to galactic latitude likely arises from residual effects of reddening on the photometry, which would become more notable at lower galactic latitudes.
We fit the trends in the top panels of Figure~\ref{fig:spatial_offset} with the displayed linear equations and subtracted the trends from our metallicities to account for these systematic effects.

We then investigated the behavior of the metallicity residuals as a function of the metallicity-sensitive color $v-g$ to calibrate for possible imperfections in our grid of synthetic photometry.
We used LAMOST, GALAH, and APOGEE metallicities as comparison samples to investigate such effects.
We find that the residuals of the photometric metallicities with respect to metallicities in those surveys show systematic trends with respect to the $v-g$ color (see Figure~\ref{fig:vg_processing}), suggesting slight imperfections in our method of deriving metallicities.
We alleviated this imperfection by reducing our photometric surface gravities by 0.55\,dex before deriving final photometric metallicities.
This adjustment to the surface gravities reduces these systematic trends (see bottom panels of Figure~\ref{fig:vg_processing}) and also brings our surface gravity scale in agreement to that used in large high-resolution spectroscopic studies of metal-poor stars (e.g., \citealt{erf+20}).

\begin{figure}[t!]
\includegraphics[width =\columnwidth]{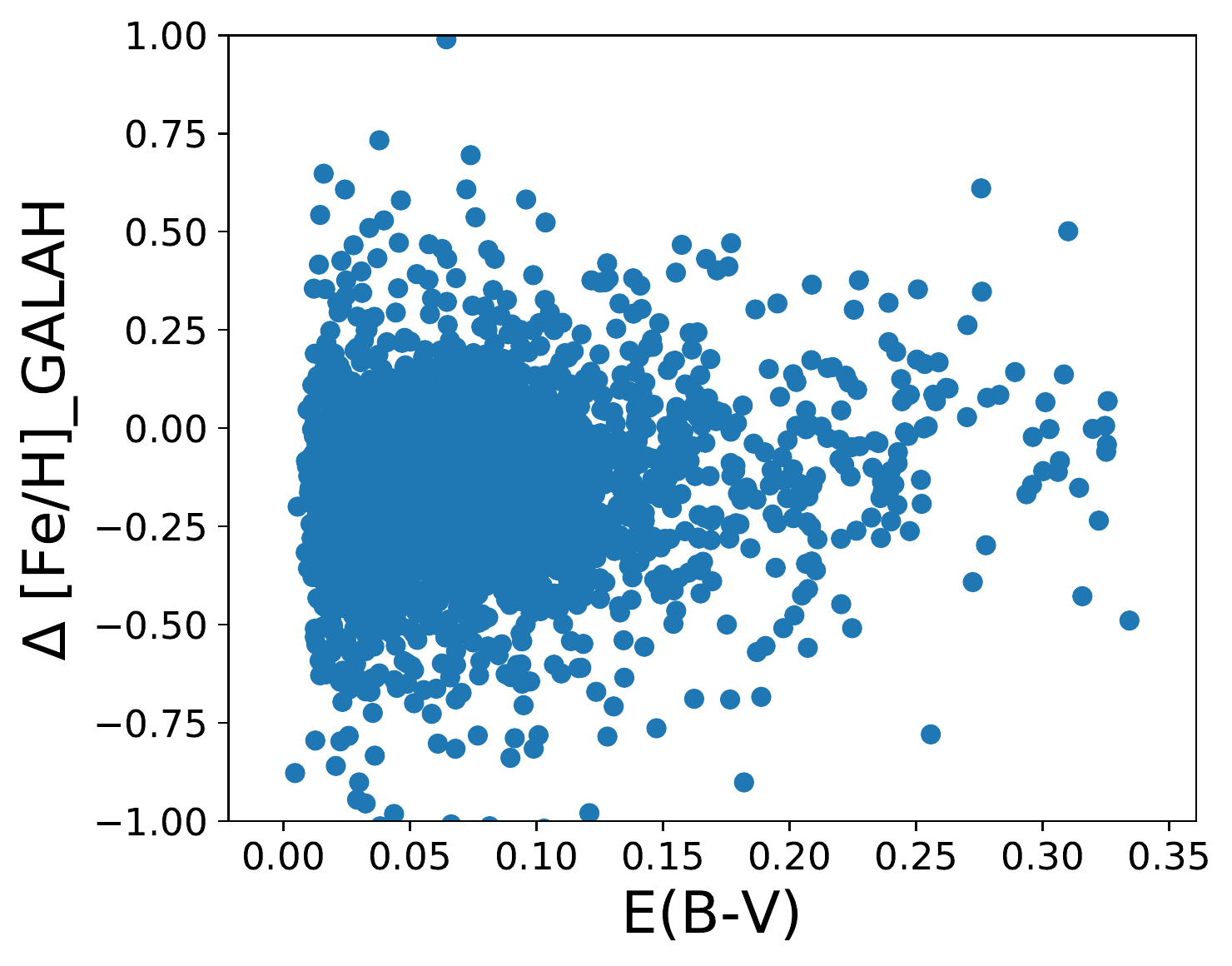}
\caption{Residuals of the final photometric metallicities with respect to metallicities from GALAH DR3 shown as a function of the reddening values from \citet{sfd+98}. 
No strong systematics appear as a function of reddening value out to E(B$-$V) $\sim0.35$, beyond an offset with metallicities in GALAH DR3.}
\label{fig:reddening_trend}
\end{figure}

We further find no strong systematics in the residuals of our photometric metallicities with respect to metallicities from GALAH DR3 as a function of reddening values from \citet{sfd+98} out to E(B$-$V) $\sim 0.35$. This is shown in Figure~\ref{fig:reddening_trend}.

\subsection{Distance-based pruning using \textit{Gaia} EDR3}
\label{sec:gaia}
We performed additional tests using \textit{Gaia} EDR3 data \citep{gaia+16, gaia+20} to flag two cases of contaminants in our catalog.
First, while we aim for our catalog to be limited to giant stars by limiting to photometric surface gravities $\log\,g < 3.0$, there may still be main-sequence interlopers due to e.g., stars with high uncertainties on their $\log\,g$ values.
Second, given the large number of stars in the Milky Way in the metallicity regime right above our sample ($-0.75 < $ [Fe/H] $< 0.0$), it is possible for some stars in that metallicity regime to contaminate the more metal-rich end ([Fe/H] $\gtrsim -1.25$) of our sample.
We thus performed two checks to identify such stars in our catalog.

\begin{figure}[t!]
\includegraphics[width =\columnwidth]{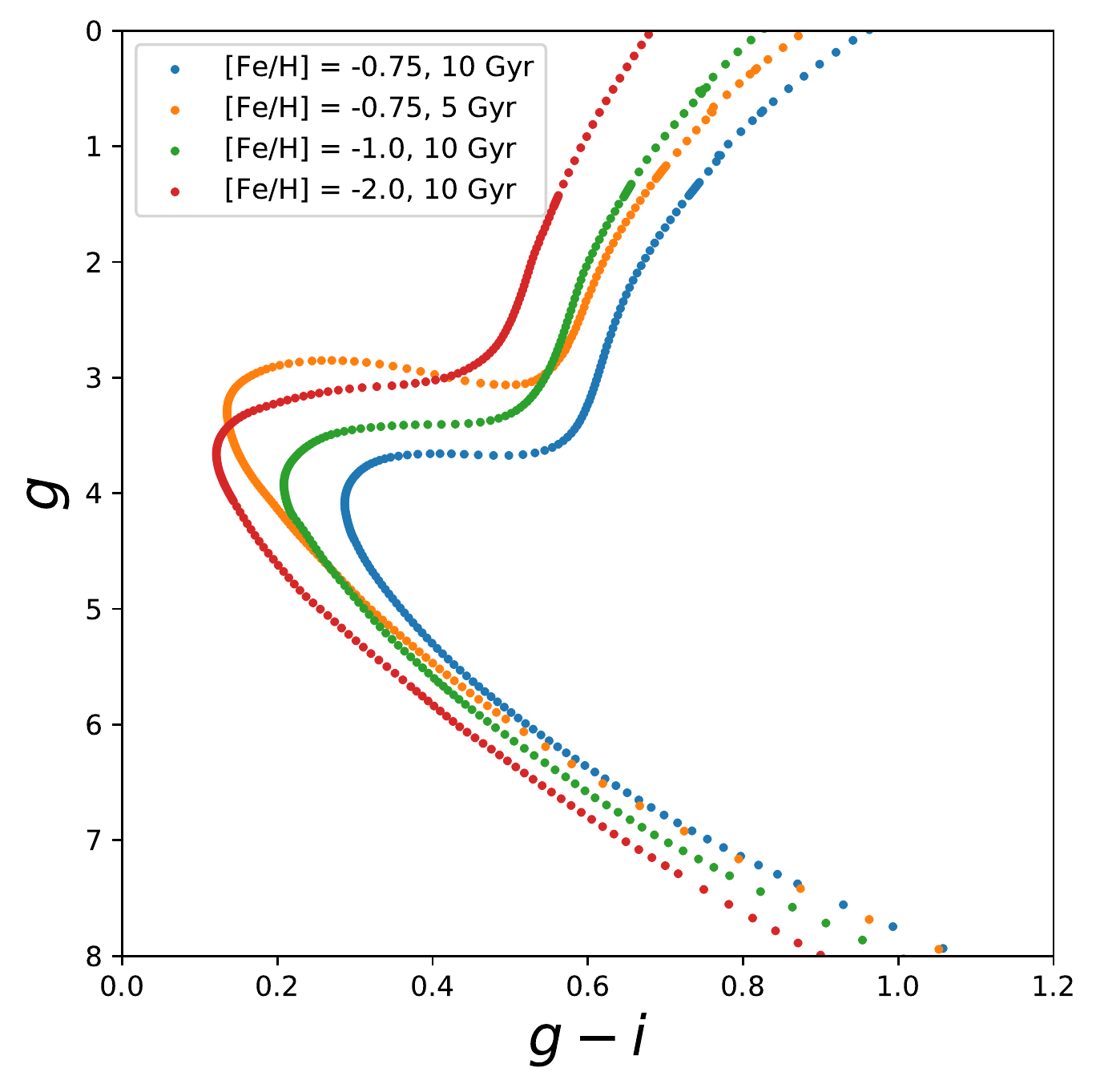}
\caption{Dartmouth isochrones of [Fe/H] $= -0.75, -1.0,$ and $-2.0$ plotted with SkyMapper $g-i$ color and absolute $g$ magnitude. 
We flag stars in our sample that have absolute SkyMapper $g > 5.0$ or that are redward of the [Fe/H] $ = -0.75$, 10\,Gyr isochrone as main-sequence or metal-rich contaminants in our sample (see Section~\ref{sec:gaia}). 
We show the [Fe/H] = $-$0.75 isochrone with two ages for illustrative purposes.}
\label{fig:isochrones}
\end{figure}

We first cross-matched our sample with \citet{bailer-jones21} to compile their photogeometric distances.
We then derived a distance modulus and absolute magnitude for each star from these distances to generate a color-magnitude diagram of our entire sample of stars. 
We identified stars that had a 84\% percentile value in their distance posterior that led to absolute SkyMapper $g$ magnitude $< 5.0$ as plausible main sequence interlopers, since such magnitudes easily exclude stars from the giant branch (see Figure~\ref{fig:isochrones}). 
Further, we then overlaid a Dartmouth isochrone \citep{dcj+08} with [Fe/H] $= -0.75$, 10\,Gyr on our color-magnitude diagram to identify plausible metal-rich stars in our catalog. 
We identified stars that had a 84\% percentile in their distance posterior that placed them red-ward of the isochrone and that had photometric [Fe/H] $> -2.0$ as plausible metal-rich interlopers.
Stars that satisfy either of these criteria have \texttt{flag\_msmr = 1} in our catalog (see Table~\ref{tab:phot_met}).

\subsection{Description of final metallicity catalog}
\begin{figure*}[t!]
\includegraphics[width =\textwidth]{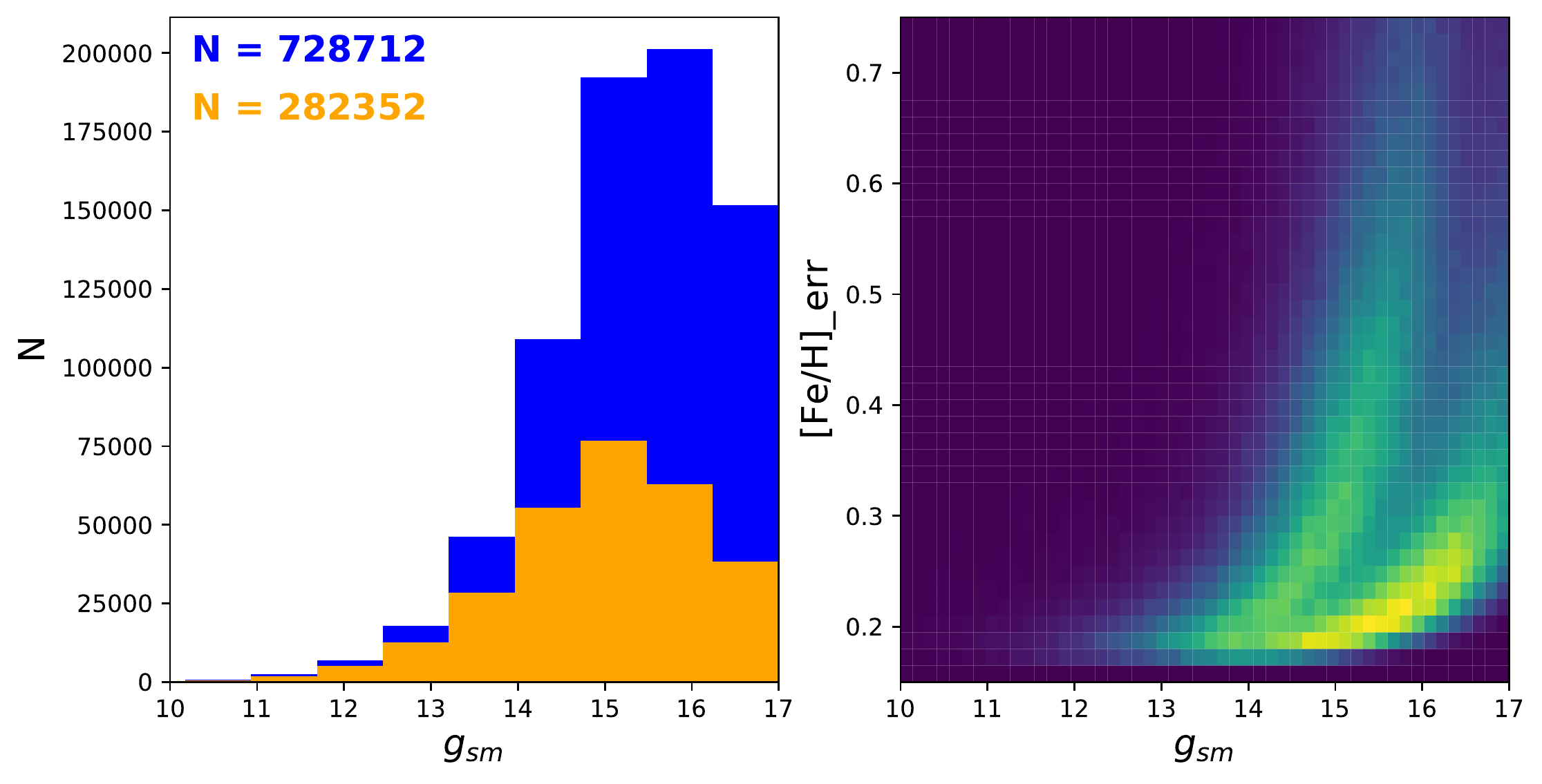}
\caption{Left: Histogram of the magnitudes of all stars in our catalog (blue) and those passing the flags in Section~\ref{sec:gaia} (orange). 
Right: Uncertainties on the photometric metallicities of the stars in our sample as a function of magnitude.}
\label{fig:summary}
\end{figure*}

Our full catalog of photometric metallicities is composed of 728,712 stars, after the implementation of a number of cuts for quality control purposes that we briefly describe here. 
First, we applied an initial photometric quality cut, as outlined in Section~\ref{sec:compile}.
Second, to ensure a basic quality of our compiled photometric metallicities, we excluded stars with a random uncertainty in their photometric metallicity (e.g., the propagated uncertainty from the photometry) of $\geq 0.75$\,dex.
Third, we applied cuts discussed in Section~\ref{sec:derive} to avoid keeping stars with stellar parameters near the edge of our grid of photometry.
Fourth, we excluded stars at galactic latitudes $|b| < 10^{\circ}$ or with E(B$-$V) $>$ 0.35 to excise regions of extreme reddening.
This ultimately resulted in our final sample of stars, which we refer to as our catalog in the following discussions.
The distribution of magnitudes and metallicity uncertainties of our catalog are shown in Figure~\ref{fig:summary}.
Our catalog is shown in Table~\ref{tab:phot_met}, which is published in its entirety in machine-readable format.
Of the 728,712 stars in our catalog, 282,351 stars pass the checks in Section~\ref{sec:gaia}.


\begin{deluxetable*}{lccccccc} 
\tablecaption{\label{tab:phot_met} Photometric metallicities of stars in SkyMapper DR2}
\tablehead{   
  \colhead{RA (deg)} & 
  \colhead{DEC (deg)} &
  \colhead{$g$ } &
  \colhead{$i$ } &
  \colhead{$\text{[Fe/H]}$} & 
  \colhead{$\sigma_{\text{[Fe/H]}}$ } & 
  \colhead{\textit{Gaia} EDR3 source ID} &
  \colhead{flag$_{\text{msmr}}$}\\
   \colhead{(J2000)}&
   \colhead{(J2000)}&
   \colhead{[mag]}&
   \colhead{[mag]}&
   \colhead{[dex]}&
   \colhead{[dex]}&
   \colhead{}&
   \colhead{}
  }
\startdata
10.46886 & $-$12.648856 & 10.35 & 9.55 & $-$1.05 & 0.22 & 2376821329509745152 & 1.0 \\
11.33971 & $-$13.854287 & 10.38 & 9.60 & $-$1.30 & 0.25 & 2375707631605778560 & 0.0\\
12.850476 & $-$10.422958 & 10.38 & 9.63 & $-$1.12 & 0.19 & 2473618420505263232 & 1.0\\
12.709095 & $-$13.133477 & 10.73 & 9.97 & $-$1.39 & 0.26 & 2376214300308134784 & 0.0\\
12.501912 & $-$13.469329 & 10.83 & 10.32 & $-$0.97 & 0.42 & 2372823990562608256 & 1.0
\enddata
\tablecomments{Table~\ref{tab:phot_met} is published in its entirety in the machine-readable format.
      A portion is shown here for guidance regarding its form and content.}

\end{deluxetable*}


\section{Validation}
\label{sec:validation}

\subsection{Comparison to Large Spectroscopic surveys}
\label{sec:surveys}

\begin{figure*}[hbtp!]
\centering
\includegraphics[width =\textwidth]{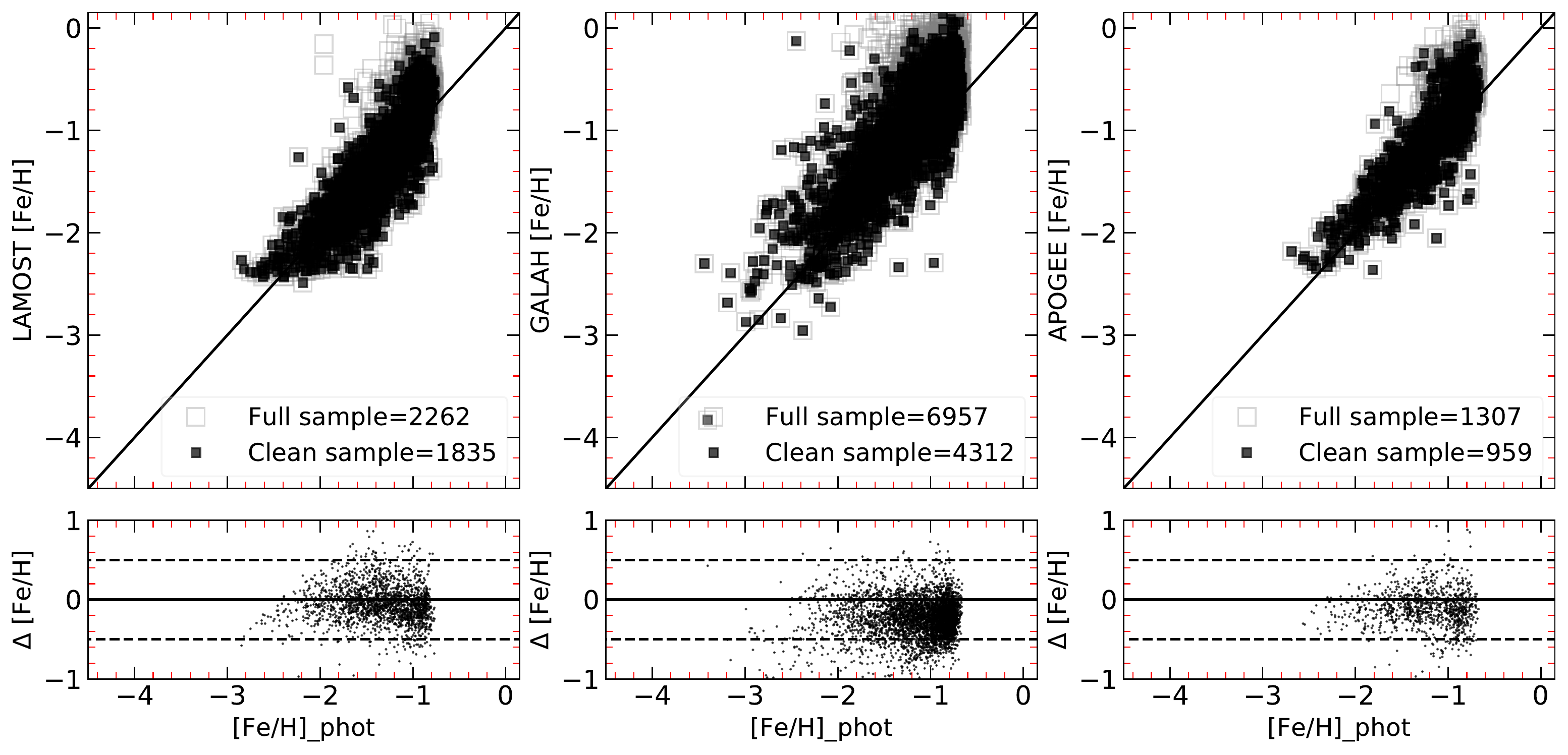}
\caption{Our SkyMapper photometric metallicities, compared to metallicities from LAMOST DR6 (left panels), GALAH DR3 (middle panels), and APOGEE DR16 (right panels) for stars with photometric metallicity uncertainties lower than 0.30\,dex. 
The filled black squares correspond to stars that passed the flags in Section~\ref{sec:gaia}, and the open squares correspond to the entire sample. 
The Lower panels show the residuals between our photometric metallicities and those presented in the above surveys.
Dashed lines indicate $\pm$0.50 to guide the eye.
The agreement between our metallicities and those in presented in these surveys is good, as indicated by the relatively low standard deviation of the residuals in the bottom panel ($\sim 0.25$\,dex) despite slight offsets with each survey.}
\label{fig:lamost}
\end{figure*}

Due to the large size of our dataset and sky coverage, there is significant overlap with several large spectroscopic surveys. To test the validity of our metallicities, for stars in common, we compared our photometric metallicities with metallicities presented in LAMOST DR6 \citep{lamost_1ref,lamost_2ref,lamost_3ref}, GALAH DR3 \citep{galahdr3}, and APOGEE DR16 \citep{ewa+11, bba+17, msf+17}. Data from these surveys were downloaded\footnote{http://dr6.lamost.org}\textsuperscript{,}\footnote{https://docs.datacentral.org.au/galah/}\textsuperscript{,}\footnote{https://www.sdss.org/dr16/irspec/spectro\_data} and cross-matched to our catalog using the \texttt{TOPCAT} software \citep{t+05}.
The results of these comparisons are presented in Figure~\ref{fig:lamost}, and are discussed further in this section.

We find 6,409 stars in our catalog that have stellar parameters in LAMOST DR6. 
We applied several cuts to the LAMOST catalog ($\logg  < 3.0$, and $4000 < \teff < 5700$) to ensure the stellar parameters of stars in the LAMOST sample were comparable to those of stars in our catalog.
Then, we excluded stars in LAMOST and our catalog that have metallicity uncertainties greater than 0.3\,dex, and stars in LAMOST with $\log\,g$ uncertainties $>0.30$ to compare stars with high-quality stellar parameters.
This resulted in a sample of 2,262 stars, which are plotted in the leftmost panel of Figure~\ref{fig:lamost} as hollow squares. 
Of those stars, 1,835 pass the \textit{Gaia}-based quality checks in Section~\ref{sec:gaia} and are plotted as solid squares.
We find that our metallicities that pass the quality checks are on average 0.05\,dex lower than the LAMOST metallicities, and the standard deviation of the residuals between metallicities is 0.24\,dex. 
This is great agreement, given typical best metallicity precisions achievable with medium-resolution spectroscopy are $\sim$ 0.15\,dex, and the systematic floor on the metallicity precision from our photometric method is $\sim$0.16\,dex.

There are 14,362 stars in our catalog that have stellar parameters in GALAH DR3.
We applied the same cuts to the GALAH catalog that we applied to the LAMOST sample to ensure a high quality stellar parameter comparison.
We also applied some cuts using flags in the GALAH catalog ({\fontfamily{qcr}\selectfont flag\_sp = 0, flag\_fe\_h == 0}) to further increase the quality of the comparison sample.
This resulted in a sample of 6,957 stars, of which 4,312 passed all the checks outlined in Section~\ref{sec:gaia}.
The comparison between the metallicities is plotted in the middle panel of Figure~\ref{fig:lamost}.
We find that these metallicities
are on average 0.24\,dex lower than those metallicities presented in GALAH and the standard deviation of the residuals between our metallicities is 0.25\,dex.

The same cuts were applied to the APOGEE DR16 catalog as for the previous catalogs to ensure a high quality reference sample.
Additionally, we only retained APOGEE stars with \texttt{ASCAPFLAG = 0} which indicated no warnings or errors were raised by the APOGEE stellar parameter pipeline \citep{gah+16}. 
This led a sample of 1,307 stars in total and 959 stars passing the checks in Section~\ref{sec:gaia}, all of which are plotted in the right panel of Figure~\ref{fig:lamost}.
Our photometric metallicities that pass the quality checks are on average 0.11\,dex lower than the APOGEE metallicities, and the residuals of metallicities between the two datasets have a standard deviation of 0.23\,dex. 

\subsection{Comparison to high-resolution spectroscopic samples}
\label{sec:hires}

There are a number of stars in our catalog that have metallicities previously determined by high-resolution spectroscopic studies. 
Metallicity comparisons are particularly useful to test the behavior of our photometric values in the very low metallicity regime ($-3.5 < $[Fe/H]$ < -2.5$). 
We thus compared our metallicities to those derived by the high-resolution studies of \citet{bcb+05}, \citet{jkf+15}, \citet{mda+19}, and \citet{erf+20} that are within the metallicity range of our grid ([Fe/H] $> -4.0$), as well as a sample of stars that was specifically observed for comparison purposes with the high-resolution MIKE spectrograph \citep{bsg+03} on the Magellan/Clay Telescope in January, October, and December 2020.
Detailed results will be presented in Ou et al., in prep. but we compare to their derived metallicities here.
For completeness, we refer the reader to Section 3 of \citet{erf+20} which comprehensively details the methodology (e.g., linelist, analysis software) used in Ou et al., in prep for deriving stellar metallicities.

\begin{figure*}[hbtp!]
\centering
\includegraphics[width =\textwidth]{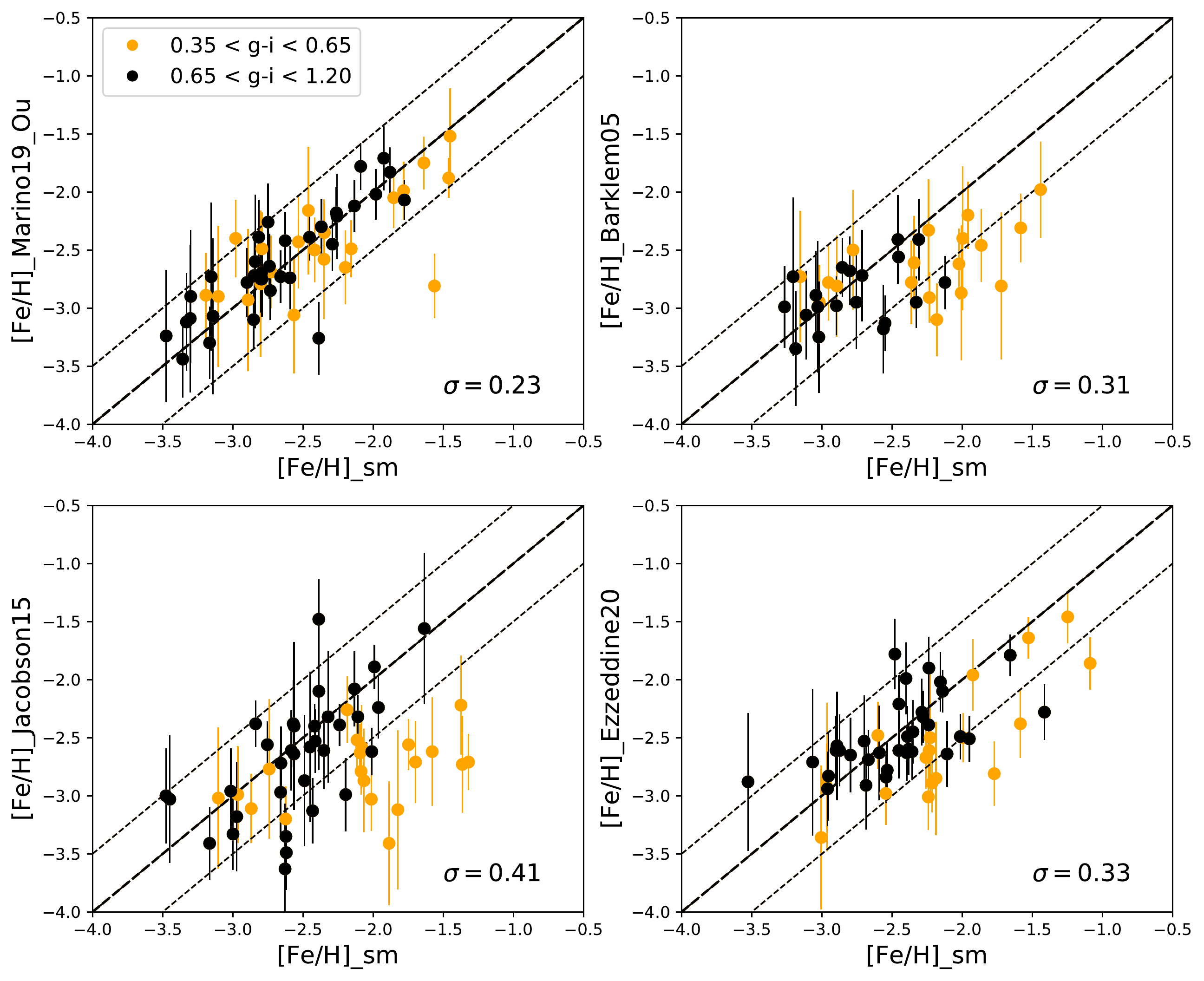}
\caption{Top left: 
A comparison between the photometric metallicities of stars presented in our catalog and metallicities from high-resolution spectroscopy of \citet{mda+19} and X. Ou et al. (in prep).
Top right: Same comparison but with the high-resolution spectroscopy results of \citet{bcb+05}.
Bottom right: Same comparison but with the high-resolution spectroscopy results of \citet{erf+20}.
Bottom left: Same comparison but with the high-resolution spectroscopy results of \citet{jkf+15}.
The black data points correspond to cooler stars in our sample ($g-i > 0.65$) and orange data points to warmer stars ($g-i < 0.65$). 
There is some evidence that our photometric metallicities are biased high with respect to three of the studies when $g-i < 0.65$.
Lines are drawn at $\pm0.5$\,dex to guide the eye. The $\sigma$ in each panel corresponds to the standard deviation of the metallicity residuals between our catalog and the corresponding study for stars with $g-i > 0.65$.}
\label{fig:hires}
\end{figure*}

Results of the various metallicity comparisons are shown in Figure~\ref{fig:hires}. The top left panel displays the comparisons to both Ou et al., (in prep) and \citet{mda+19}. 
The following panels, in clockwise order, show the comparisons to \citet{bcb+05}, \citet{erf+20}, and \citet{jkf+15}, respectively.
The orange data points in the panels correspond to the warmer stars in our sample ($g-i < 0.65$) and the black data points correspond to the cooler stars ($g-i > 0.65$). 
For reference, $g-i = 0.65$ corresponds to an effective temperature of $\sim5000$\,K.

The agreement of photometric metallicities for all 55 stars in common with those of the combined sample of \citet{mda+19} and Ou et al. (in prep) is excellent. 
Our metallicities are marginally higher (0.15\,dex) on average but the standard deviation of the residuals is 0.29\,dex.
Note that in the Figure~\ref{fig:hires} we list a slightly different standard deviation that refers to stars with $g-i > 0.65$.

We find worse agreement (mean offset of 0.24\,dex with a standard deviation of 0.45\,dex) for the metallicities of our stars in common with the three other studies \citep{bcb+05, jkf+15, erf+20} combined but there are hints that this is the case because our photometric metallicities are biased high when $g-i < 0.65$.
If we only include the 87 cooler stars with $g-i > 0.65$, the average offset drastically reduces to 0.08\,dex, and the standard deviation of the residuals reduces to 0.36\,dex. 
Finally, combining all five high-resolution samples and restricting the comparison to $-4.0 < $ [Fe/H] $< -2.5$ (as measured on our metallicity scale), metallicities agree reasonably well mean offset of 0.03\,dex with a standard deviation of 0.30\,dex), with roughly similar scatter when $g-i > 0.65$ (mean offset on 0.02\,dex with a standard deviation of 0.31\,dex).

To further gauge precision in the lowest metallicity regime ($-3.5 \lesssim$ [Fe/H] $\lesssim -2.5$), we investigated the contamination rate of stars with [Fe/H] $< -4.0$ in our catalog by cross-matching our catalog to the sample of ultra metal-poor stars listed in \citet{efp+17}.
We recover only three of their sixteen stars: CD\,$-38^{\circ}$\,245, HE~2139$-$5432, and SMSSJ0313$-$6708, suggesting a negligible presence of stars below [Fe/H] $< -4.0$ in our catalog.
Furthermore, the presence of HE~2139$-$5431 and SMSSJ0313-6708 in our catalog is not altogether surprising, given their extremely large relative carbon abundances ([C/Fe] = 2.59; \citealt{ynb+13} and [C/Fe] $> 4.8$; \citealt{kbf+14}, respectively) which significantly up-scatters their photometric metallicities (see Section~\ref{sec:carbon}).
Our recovery of CD\,$-38^{\circ}$\,245 ([Fe/H] = $-4.12\pm0.10$; \citealt{erf+20}) cannot be explained this way as it has no strong overabundance of [C/Fe], but we note a [Fe/H] $=-3.05 \pm 0.48$ for this star in our catalog. 
In general, anomalously high or low [$\alpha$/Fe] values, unresolved interstellar Ca absorption, significant CH absorption or other issues may mask as stellar Ca, and are known to result in systematic overestimates for stars when the signal from the Ca II K line has become incredibly weak (e.g., when [Fe/H] $\lesssim -3.5$). 
This highlights the difficulty in extending this technique to derive precise metallicities at the very lowest [Fe/H] regime (e.g, ultra metal-poor stars).

This comparison exercise with results from high-resolution spectroscopy principally validates our analysis technique and confirms that our photometric metallicities are reliable down to $\mbox{[Fe/H]}\sim-3.3$ for the cooler subsample of our catalog ($g-i > 0.65$). 
It also suggests that the photometric metallicities presented in this catalog could feasibly be used for targeted searches for the most metal-poor stars in our galaxy, especially among the cooler stars.
The use of metallicities of warmer stars with $g-i < 0.65$ is less accurate due to some evidence of systematic metallicity offsets when compared to \citet{bcb+05}, \citet{jkf+15}, and  \citet{erf+20},  but we still report their metallicities because (1) of the overall good agreement with the Ou et al. (in prep) and \citet{mda+19} studies, and (2) even if these metallicities were somewhat biased high, they would still be useful for targeted spectroscopic follow-up campaigns for low metallicity stars.

\subsection{Effect of carbon abundance}
\label{sec:carbon}
As extensively discussed in e.g., \citet{dbm+19} and \citet{cfj+20}, stars with a prominent CN spectral feature at $\sim$3870\,{\AA} may have photometric metallicities from SkyMapper photometry that are biased high. 
This is because that absorption feature is located within the wavelength region covered by the SkyMapper $v$ filter and thus affects the observed flux and will make the star appear more metal-rich. 
Such an effect could be particularly pronounced at lower metallicities where carbon-rich stars become more frequent.

At face value, this effect implies that carbon-enhanced metal-poor (CEMP) stars in this catalog will appear to have higher metallicities than their true metallicity.
For reference, a star with \teff = 4700\,K, [Fe/H] = $-2.5$, and [C/Fe] = 0.50 will have a metallicity biased upward by $\sim0.40$\,dex compared with a similar star with [C/Fe] = 0 \citep{cfj+20}.
This means that for the majority of stars in our catalog, this effect will be below that $\sim0.40$\,dex level since most stars with $-3.0 < $ [Fe/H]$ < -1.0$ do not reach that level of carbon enhancement (Figure~2 in \citealt{pfb+14}). 
Obvious exceptions to this would be e.g., CEMP-s stars (typically with [Fe/H] $\gtrsim -2.5$,) that display extreme carbon enhancements due to mass transfer from a binary companion but those numbers are expected to be comparably small. 
We also emphasize that if variations in the carbon abundance were greatly influencing our metallicity determinations for a bulk of stars in this catalog, this effect would have manifested as a much larger scatter in our comparisons to the metallicities obtained by the LAMOST, GALAH, and APOGEE surveys (see also Section~\ref{sec:surveys}) which is not observed. 
Hence, for the bulk of stars in this catalog, we regard the effect of the carbon abundance on the metallicity as likely not significant.

\subsection{Selection Effects on the Metallicity Distribution}

We also attempt to quantify the extent to which our catalog is biased in selecting or de-selecting stars as a function of metallicity.
We test for any such effect among stars with $-2.0 <$ [Fe/H] $< -1.0$ by deriving the fraction of stars that we recover in the southern hemisphere coverage of the LAMOST DR6, GALAH DR3 and APOGEE DR16 catalogs with comparable stellar parameters ($\logg < 3.0$, $-2.50 < $[Fe/H]$ < -0.75$, 4000\,K $<$ \teff $< 5700$\,K, and $\sigma_{\text{[Fe/H]}} < 0.3$) and spatial distribution ($|b| > 10^{\circ}$), over a magnitude range $11 \lesssim g \lesssim 16.5$.

We note that our recovery rate with respect to the combined sample of LAMOST, GALAH, and APOGEE stars is between 40\,\% and 50,\%, but this apparently low value is largely driven by our initial photometric quality cut on the SkyMapper DR2 catalog (\texttt{flags = 0}).
After that initial cut, a significantly higher 70\,\% of stars in the LAMOST, GALAH, and APOGEE catalogs are retained in our photometric metallicity catalog, suggesting that the majority of the loss is driven by photometric quality and not anything particular to our subsequent analysis technique.
However, we still present the recovery fraction below including the loss of stars in our catalog due to photometric quality, in order to test whether such losses could plausibly have an effect on the metallicity selection.
Since SkyMapper imaging covers the entire southern sky, we note that our catalog includes a significant number of stars that are in neither the LAMOST, GALAH, or APOGEE catalogs.

We determined that for $-1.5 < $[Fe/H]$ < -1.0$, 42\,\% of stars were retained in our catalog. 
For $-2.0 < $[Fe/H]$ < -1.5$, 47\,\% were retained. 
At face value, this suggests that we are slightly more sensitive to accurately selecting stars with $-2.0 < $[Fe/H]$ < -1.5$ than  $-1.5 < $[Fe/H]$ < -1.0$. This difference in recovered stars is robust given the large sample size of 8270 reference stars from LAMOST, GALAH and APOGEE.

To test whether any potential selection effects extend into the [Fe/H] $< -2.0$ regime, we performed the same exercise, except this time using the combined sample of stars from \citet{bcb+05}, \citet{jkf+15}, \citet{mda+19}, \citet{erf+20}, and X. Ou et al. (in prep) as references, as they extend to lower metallicities than the APOGEE/LAMOST$/GALAH$ comparison.
In this case, we find no strong dependency of the recovery fraction with metallicity, with recovery fractions of 58\,\%, 51\,\%, \,45\%, and 25\,\% for the respective 0.5\,dex increments ranging from [Fe/H] $= -1.5$ to [Fe/H] = $-3.5$. 
Unfortunately, the smaller sample size of 417 stars in this combined sample is too small to resolve differences below a $\sim7\,\%$ level (set by Poisson statistics) in each of these bins.
For that reason, it is particularly difficult to asses the completeness in the extremely metal-poor regime with [Fe/H] $< -3.0$ at this time, although indications suggest that our catalog is preferentially incomplete in that regime. 
This is not surprising, as the selection is likely to decrease due to a loss of metallicity precision since the color-color separation between stars of different metallicities significantly narrows in this regime (see Figure~\ref{fig:photometric_metallicity}).
We note that including/excluding stars that fail the quality checks in Section~\ref{sec:gaia} negligibly changes these recovery fractions.

\section{Summary and Conclusion} \label{sec:conclusion}

In this paper, we have presented a new catalog of $\sim720,000$ stars in the Southern hemisphere for which we have obtained photometric metallicities using  metallicity-sensitive photometry from the second data release (DR2) of the SkyMapper Southern Survey. 
We identify $\sim280,000$ of these stars as having reliable metallicities, after excluding main-sequence and more metal-rich ([Fe/H] $> -0.75$) contaminants using \textit{Gaia} EDR3 data.
This sample of giants with T$_{\rm eff}\lesssim$ 5600\,K, $\log g < 3.0$, and $-3.75 <$ [Fe/H] $< -0.75$ reaches down to $g=17$, and stretches throughout the inner halo to a scale height of $|Z|\sim 7$\,kpc (Chiti et al. ApJL accepted).
We find that our photometric metallicities compare well (average offsets $\sim 0.15$\,dex with standard deviation $\sigma \sim 0.25$\,dex) to those obtained in large-scale surveys (e.g., LAMOST, GALAH, APOGEE), suggesting that our analysis techniques produce accurate metallicities for the bulk of the stars in our sample.
Overall, this validates our general approach of deriving photometric metallicities.

To test the performance of our catalog in the [Fe/H] $< -2.5$ regime, we compared our photometric metallicities to metallicities derived in several high-resolution spectroscopic studies of very and extremely metal-poor stars in \citet{bcb+05}, \citet{jkf+15}, \citet{mda+19}, \citet{erf+20}, and X. Ou et al. (in prep). 
We find some evidence that warmer stars in our catalog ($g-i < 0.65$) have systematically higher photometric metallicities. 
However, we find good agreement (mean offset = 0.02\,dex with standard deviation $\sigma = 0.31$\,dex) between the photometric metallicities of cooler stars with $g-i > 0.65$ with [Fe/H] $< -2.5$ and their metallicities from high-resolution studies. 

We highlight that several systematic effects could bias our photometric metallicities. 
First, as noted before, warmer stars with $g-i < 0.65$ may have metallicities biased high, at least when compared to several high-resolution studies of very and extremely metal-poor stars.
Secondly, carbon-enhanced metal-poor (CEMP) stars will systematically have artificially higher photometric metallicities due to the presence of a CN absorption feature in the bandpass of the SkyMapper $v$ filter.
However, the latter effect likely does not affect the bulk of stars in our sample as evidenced by the good agreement of our photometric metallicities to metallicities presented in APOGEE and LAMOST. 

In Chiti et al. (ApJL accepted), we use a subset of this catalog to explore the metallicity distribution function (MDF) of the Milky Way and create spatial metallicity maps of our galaxy. 
We recover well-known features in the MDF, such as a peak at [Fe/H] $\sim -1.5$ when considering stars distant from the disk plane ($|Z| > 5$\,kpc) \citep{isj+08}.
We also find that the metallicity distribution function steepens below [Fe/H] = $-2.3$, confirming the significant challenge in searching for the most metal-poor stars, and we identify of order 1000 giants with quality photometric metallicities (random uncertainties $< 0.50$\,dex) and colors ($g-i > 0.65$) that have [Fe/H] $\lesssim -2.6$ in this catalog. 

This shows that this catalogue is suitable for a variety of chemical characterizations of the metal-poor Galaxy ([Fe/H] $<-0.75$) and its components, as well as targeted searches of stellar populations such as the most metal-poor stars. Accordingly, spectroscopic observations and detailed kinematic analyses are currently underway to further characterize the low-metallicity tail of the metallicity distribution function and to obtain detailed chemical abundances of the population of stars with [Fe/H] $<-3.0$. This will contribute to our understanding of the origin and evolution of the oldest components of the Milky Way.

\software{Astropy \citep{astropy}, NumPy \citep{numpy}, SciPy \citep{scipy}, Matplotlib \citep{Hunter+07}}

\acknowledgements
A.C. and A.F. acknowledge support from NSF grant AST-1716251. A.C thanks Dougal Mackey for advice on de-reddening the SkyMapper photometry. A.F. thanks the Wissenschaftskolleg zu Berlin for their wonderful Fellow's program and hospitality. 
This work made use of NASA's Astrophysics Data System Bibliographic Services, and the SIMBAD database, operated at CDS, Strasbourg, France \citep{woe+00}.

This work has made use of data from the European Space Agency (ESA) mission
{\it Gaia} (\url{https://www.cosmos.esa.int/gaia}), processed by the {\it Gaia}
Data Processing and Analysis Consortium (DPAC,
\url{https://www.cosmos.esa.int/web/gaia/dpac/consortium}). Funding for the DPAC
has been provided by national institutions, in particular the institutions
participating in the {\it Gaia} Multilateral Agreement.

The national facility capability for SkyMapper has been funded through ARC LIEF grant LE130100104 from the Australian Research Council, awarded to the University of Sydney, the Australian National University, Swinburne University of Technology, the University of Queensland, the University of Western Australia, the University of Melbourne, Curtin University of Technology, Monash University and the Australian Astronomical Observatory. SkyMapper is owned and operated by The Australian National University's Research School of Astronomy and Astrophysics. The survey data were processed and provided by the SkyMapper Team at ANU. The SkyMapper node of the All-Sky Virtual Observatory (ASVO) is hosted at the National Computational Infrastructure (NCI). Development and support the SkyMapper node of the ASVO has been funded in part by Astronomy Australia Limited (AAL) and the Australian Government through the Commonwealth's Education Investment Fund (EIF) and National Collaborative Research Infrastructure Strategy (NCRIS), particularly the National eResearch Collaboration Tools and Resources (NeCTAR) and the Australian National Data Service Projects (ANDS).

Funding for the Sloan Digital Sky Survey IV has been provided by the Alfred P. Sloan Foundation, the U.S. Department of Energy Office of Science, and the Participating Institutions. SDSS acknowledges support and resources from the Center for High-Performance Computing at the University of Utah. The SDSS web site is www.sdss.org.

SDSS is managed by the Astrophysical Research Consortium for the Participating Institutions of the SDSS Collaboration including the Brazilian Participation Group, the Carnegie Institution for Science, Carnegie Mellon University, Center for Astrophysics | Harvard \& Smithsonian (CfA), the Chilean Participation Group, the French Participation Group, Instituto de Astrofísica de Canarias, The Johns Hopkins University, Kavli Institute for the Physics and Mathematics of the Universe (IPMU) / University of Tokyo, the Korean Participation Group, Lawrence Berkeley National Laboratory, Leibniz Institut für Astrophysik Potsdam (AIP), Max-Planck-Institut für Astronomie (MPIA Heidelberg), Max-Planck-Institut für Astrophysik (MPA Garching), Max-Planck-Institut für Extraterrestrische Physik (MPE), National Astronomical Observatories of China, New Mexico State University, New York University, University of Notre Dame, Observatório Nacional / MCTI, The Ohio State University, Pennsylvania State University, Shanghai Astronomical Observatory, United Kingdom Participation Group, Universidad Nacional Autónoma de México, University of Arizona, University of Colorado Boulder, University of Oxford, University of Portsmouth, University of Utah, University of Virginia, University of Washington, University of Wisconsin, Vanderbilt University, and Yale University.

Guoshoujing Telescope (the Large Sky Area Multi-Object Fiber Spectroscopic Telescope LAMOST) is a National Major Scientific Project built by the Chinese Academy of Sciences. Funding for the project has been provided by the National Development and Reform Commission. LAMOST is operated and managed by the National Astronomical Observatories, Chinese Academy of Sciences.

\bibliography{skymapper}

\end{document}